\documentclass[3p,times,procedia]{elsarticle}
\usepackage{ecrc}
\usepackage[bookmarks=false]{hyperref}
    \hypersetup{colorlinks,
      linkcolor=blue,
      citecolor=blue,
      urlcolor=blue}
\usepackage{subcaption}
\usepackage{algpseudocode}
\usepackage{algorithm}
\usepackage{algorithmicx}
\usepackage{amsthm}

\theoremstyle{lemma}
\newtheorem{lemma}{Lemma}
\theoremstyle{theorem}
\newtheorem{theorem}{Theorem}

\newcommand{\argmin}{\mathop{\rm arg~min}\limits}

\usepackage{xspace}
\usepackage{booktabs}
\usepackage{amsmath}
\usepackage{amssymb}
\usepackage{amsfonts}
\usepackage{wrapfig}
\usepackage{color}

\def\ignore#1{}

\def\AgentSet{A}
\def\posInt{\mathbb{Z}_+}
\def\pibtFalse{\mathit{False}}
\def\pibtTrue{\mathit{True}}
\def\e{\varepsilon}
\def\exPIBT{{\sc exPIBT}\xspace}
\def\exPIBTTA{{\sc exPIBT-TA}\xspace}
\def\PIBT{{\sc PIBT}\xspace}
\def\Reserve{{\sc Reserve}\xspace}
\def\Revert{{\sc Revert}\xspace}
\def\valid{{\sc valid}\xspace}
\def\invalid{{\sc invalid}\xspace}
\def\TaskSet{\mathcal{T}}

\volume{00}

\firstpage{1}

\journalname{Procedia Computer Science}

\runauth{Y. Fujitani, T. Yamauchi, Y. Miyashita and T. Sugawara}

\jid{procs}

\usepackage{amssymb}

\usepackage[figuresright]{rotating}

\begin{document}
\begin{frontmatter}




\title{Deadlock-Free Method for Multi-Agent Pickup and Delivery Problem Using Priority Inheritance with Temporary Priority}


\author[w]{Yukita Fujitani\corref{cor1}}
\author[w]{Tomoki Yamauchi}
\author[w]{Yuki Miyashita}
\author[w]{Toshiharu Sugawara}


\address[a]{Department of Computer Science and Communications
  Engineering, Waseda University, Tokyo 1698555, Japan}


\begin{abstract}
  This paper proposes a control method for the multi-agent pickup and
  delivery problem (MAPD problem) by extending the {\em priority
    inheritance with backtracking} (PIBT) method to make it applicable
  to more general environments. PIBT is an effective algorithm that
  introduces a priority to each agent, and at each timestep, the
  agents, in descending order of priority, decide their next
  neighboring locations in the next timestep through communications
  only with the local agents. Unfortunately, PIBT is only applicable
  to environments that are modeled as a bi-connected area, and if it
  contains dead-ends, such as tree-shaped paths, PIBT may cause
  deadlocks. However, in the real-world environment, there are many
  dead-end paths to locations such as the shelves where materials are
  stored as well as loading/unloading locations to transportation
  trucks. Our proposed method enables MAPD tasks to be performed in
  environments with some tree-shaped paths without deadlock while
  preserving the PIBT feature; it does this by allowing the agents to
  have temporary priorities and restricting agents' movements in the
  trees. First, we demonstrate that agents can always reach their
  delivery without deadlock. Our experiments indicate that the
  proposed method is very efficient, even in environments where PIBT
  is not applicable, by comparing them with those obtained using the
  well-known {\em token passing} method as a baseline.
\end{abstract}


\cortext[cor1]{Corresponding author.}
\end{frontmatter}

\email{y.fujitani@isl.cs.waseda.ac.jp}




\section{Introduction}
With the recent development of artificial intelligence technology, intelligent agents, which are models of machines or systems that can recognize their environment and autonomously act accordingly, have attracted recent attention, and have thus been extensively research for use in various applications. For example, a cleaning robot (agent) can learn the layout of a room without prior information, and can automatically clean it. There are also examples of agents exploring people and objects in areas that are inaccessible to humans during a disaster. When these agents are expected to be used for complex tasks or in large environments, multiple agents are required to complete these tasks by coordinating and cooperating with each other to achieve their own goals or shared goals. There is a wide range of multi-agent system applications, for example, traffic flow control~\cite{dresner2008}, robots in an automated warehouse~\cite{wurman2008}, cooperative security surveillance~\cite{Sugiyama2019}, and airplane operation control~\cite{morris2016}. However, appropriate coordinated behavior is sophisticated, and just taking optimal actions based on agents' independent decisions may lead to conflicts with other agents, such as competition for shared and limited resources and physical collisions. Because these conflicts occur more frequently with the increase of agents and hinder the efficiency of entire systems, it is crucial to control all agents to reduce the possibility of conflicts.
\par

Among the many applications of multi-agent systems, we focus on the {\em multi-agent pickup and delivery} (MAPD) problem~\cite{ma2017,ma2019} as a fundamental problem, in which multiple agents continuously perform multiple pickup-and-delivery tasks in a particular environment in parallel without collision. Therefore, we can consider an MAPD instance as an asynchronous iteration of {\em multi-agent path finding} (MAPF) problems, in which multiple agents find collision-free paths from their current positions to their own deliveries. A task in MAPD is expressed by a pair of start and goal locations, and the agent assigned the task has to carry a material in the start location to the goal location without collision. Their aim is to work together to complete all required MAPD tasks as quickly as possible.
\par

Although several algorithms have been proposed to solve the MAPD problem~\cite{ma2017,okumura2017,cbs2015} as discussed in the next section, we focus on {\em priority inheritance with backtracking} (PIBT)~\cite{okumura2017} because it enables decentralized and collision-free continuous task execution. In PIBT, agents locally calculate their priorities at every step, and decide their next moving locations with no conflict in order of priority between agents by communicating only with neighboring agents. PIBT appears to be scalable to an increasing number of agents because each agent decides its next location locally, but the structure of the environment must be a bi-connected graph, i.e., any pair of two nodes must have a path connecting them, even if one other arbitrary node is removed\footnote{This can also be described as two arbitrary nodes having multiple paths between them that do not share a common node.}. For example, this restriction means that deadlocks may occur in environments containing short dead-end paths, such as cul-de-sacs or tree-structured paths. When considering automated robotic delivery problems in realistic warehouses, loading and unloading of racks and trucks are often performed in dead-ends, and so the naive PIBT cannot be applicable. There are other methods that can be applied in environments containing dead-ends, such as well-known {\em token passing} (TP)~\cite{ma2017}. However, TP requires the condition where agents do not pass through or set as the destination the loading/unloading locations of tasks that are currently being executed by other agents. This requirement reduces parallelism and often spoils the benefits of multi-agent systems, making it difficult to improve efficiency.
\par

Therefore, we propose two algorithms that are extensions of PIBT for
application to environments where the constraints required by PIBT are
relaxed. More specifically, considering the MAPD problems for real-world applications, such as automated carrying robots in a warehouse, pickup-and-delivery robots on a construction site, and rescue robots in a disaster situation, our proposed extended PIBT can be used without causing a deadlock in environments where a number of tree-structured dead-end paths are connected to the main area, which is a bi-connected graph as required by PIBT. Besides the priority used in the conventional PIBT, we have introduced {\em temporary priorities} into the algorithms, so that reachability between any pair of nodes that do not exist in the same tree path can be guaranteed in environments containing dead-end paths and trees, while avoiding deadlocks, without changing the favorable features of the PIBT. The first proposed algorithm is a simple base extension of PIBT, where agents travel only the shortest path between the root node and the destination node at the end of a tree. The second algorithm is its improved version for efficiency, so that if possible, an agent can wait temporarily on a side branch in the tree to allow agents to cross. We conduct comparative experiments using a number of MAPD instances in several settings with the proposed algorithms and TP~\cite{ma2017} as a baseline. The results show that in most cases, the proposed algorithms are more effective and efficient than the baseline algorithm.
\par

\section{Related Work}
There have been many studies that focus on MAPF problems to generate
collision-free paths for multiple
agents~\cite{goldenberg2014,silver2005,standley2010,wagner2015}. For
example, Silver~\cite{silver2005} proposed {\em cooperative A*} and
its extension, whereby each agent generates a collision-free path,
from information about the plans of the other agents. Wagner and
Choset~\cite{wagner2015} proposed an {\em enhanced partial expansion
  A*}, which is an efficient version of A* search, and attempted to
apply it to an MAPF problem.
\par

There are two main approaches to path generation and its control in MAPD, which is an iteration of MAPF: a centralized control method in which a specific agent grasps the entire situation and generates all plans~\cite{pushandswap,cbs2015}, and a decentralized method in which individual agents autonomously generate their own plans according to their local surroundings~\cite{alessandro2020,kinetic_ma,ma2017,okumura2017}. For example, with the former approach, Sharon et al.~\cite{cbs2015} proposed {\em conflict-based search} (CBS), which consists of two stages, high-level and low-level searches, and generates paths that are conflict-free with the already generated paths. Luna and Bekris~\cite{pushandswap} introduced two operations that involve pushing an agent closer to the goal and swapping the positions of two agents to control the movement of the multiple agents without collision. However, with centralized control methods, costs are likely to increase as the number of agents increases, which may reduce the overall efficiency.
\par

In contrast, Ma et al.~\cite{ma2017} proposed a decentralized algorithm, TP, for an environment such as Amazon's warehouse, which is pre-designed for automated delivery by robots. In this algorithm, the paths of agents currently being executed are stored in the memory shared with all agents called {\em token}, and agents with permission to access it autonomously generate new conflict-free plans, and store them in the token. Farinelli, Contini, and Zorzi~\cite{alessandro2020} extended TP to loosen the conditions required by TP, and applied it to the relaxed MAPD in which a robot can deliver multiple materials. Ma et al.~\cite{kinetic_ma} introduced another extension of TP in which physical constraints such as the size and speed of robots are considered. Yamauchi et al.~\cite{Yamauchi:aamas2022} proposed the {\em standby-based deadlock avoidance} algorithm, and by integrating it with TP, agents can achieve a high degree of parallelism in a more general environment. Although these decentralized control approaches, including PIBT~\cite{okumura2017}, have the potential to prevent increased costs because of the increased number of agents, they usually assume some constraints to restrict in terms of environment and/or task selection. Therefore, by relaxing the environmental conditions required by PIBT, we aim to expand the range of PIBT applications.

\section{Preliminary}
\subsection{Problem Description}
Let $\AgentSet=\{a_1,a_2,\dots,a_n\}$ be a set of $n$ agents. We introduce a discrete-time $t\in\posInt$ whose unit is {\em timestep} ($\posInt$ is the set of positive integers). The environment in which agents move around is denoted by a undirected graph $G=(V,E)$, which can be embedded into a two-dimensional (2D) Euclidean space. An agent can stay at node $v\in V$ and move to its neighboring node $u\in N_v$, where $N_v=\{u \mid \exists (v,u)\in E\}$. Like PIBT, we assume that the length of all edges is one, and that agents can move to a neighboring node in one timestep. Note that PIBT assumes that $G$ is bi-connected and does not contain self-loops, multiple edges, and dead-end nodes.
\par

A task $\tau$ in MAPD is expressed by $\tau=(s^\tau, g^\tau)$, where $s^\tau\in V$ is the pickup node and $g^\tau\in V$ is the delivery node. Therefore, $a_i\in\AgentSet$ assigned $\tau$ first moves to $s^\tau$ to pick up a material, moves to $g^\tau$, and then unload the carried material. For simplicity, we assume that $a_i$ loads the material when $a_i$ arrives at $s^\tau$, and when $a_i$ arrives at $g^\tau$, $a_i$ unloads it (so $a_i$ completes $\tau$). A MAPD instance, which is the set of tasks to be completed, is denoted by $\TaskSet$. When $\TaskSet$ is given, each agent is assigned a task or chooses a task one-by-one depending on the application requirements.
\par

The location (node) of agent $a_i$ at timestep $t$ is denoted by $v_i(t)\in V$. Agent $a_i$ can move to a neighboring node or can stay at the current position; thus,
\begin{center}
  $v_i(t+1) \in N_{v_i(t)} \cup \{v_i(t)\}$
\end{center}
To avoid collisions, agents cannot be on the same node and cannot pass each other, i.e.,
\begin{center}
  \(v_i(t) \neq v_j(t), \;\; \mathrm{ and }\;\;
  v_i(t) \neq v_j(t+1) \vee v_i(t+1) \neq v_j(t).\)
\end{center}
Assuming synchronized actions, the following circulation actions are possible.
\begin{center}
  \(v_i(t+1)= v_j(t) \wedge v_j(t+1)=v_k(t) \wedge \dots \wedge
  v_l(t+1)=v_i(t)\)
\end{center}
Note that $|V| \geq n$ must be satisfied.
\par

\subsection{PIBT}
Next, we briefly explain PIBT~\cite{okumura2017}. PIBT is an algorithm in which each agent $a_i\in\AgentSet$ at every time step calculates its own priority $p_i$, which is the {\em strict total order} in $\AgentSet$. It then decides the next node in order from the agent with the highest priority in the local range. To do this, PIBT assumes that (1) agent $a_i$ can decide the rank (desired order) of its next neighboring node in $N_{v_{i(t)}}\cup\{v_i(t)\}$ to move based on the path to $a_i$'s current destination $d_i$, which is the pickup or delivery node of the current task, using the map of the environment; (2) it has the map of environment $G$ and can identify its location; (3) it can communicate with other agents within a distance of two (nodes that can be reached by two edges); and (4) all agents move synchronously when the next nodes of all agents have been decided.
\par

First, all agents attempt to move to the first ranked neighboring node within the next timestep, but they also have to prevent collisions and deadlocks between agents by recursively performing {\em priority inheritance} (PI) to neighboring agents when necessary. PI means that when $a_i$ is on the node to which $a_j$ with higher priority ($p_i < p_j$) also wants to move, $a_i$ must move from that node, but at the same time inherits the priority of $a_j$ (so, $p_i\leftarrow p_j$). Furthermore, if another agent $a_k$ competes for that node by selecting it as the $a_k$'s next node, $a_i$ consequently inherits the higher priority of $a_j$ and $a_k$ ($p_i\leftarrow \max(p_j, p_k)$), and causes the other to abandon its current next node.
\par

\begin{wrapfigure}[7]{R}[0mm]{7.8cm}
  \begin{minipage}{0.32\hsize}
    \centering
    \includegraphics[width=0.85\hsize]{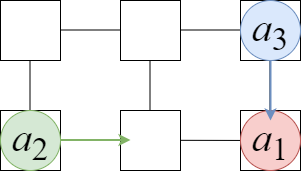}
    \subcaption{Stalemate}
    \label{subfig:ikidumari}
  \end{minipage}
  \hfil
  \begin{minipage}{0.32\hsize}
    \centering
    \includegraphics[width=0.85\hsize]{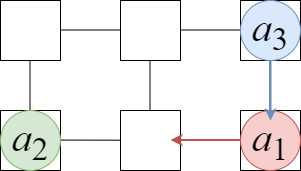}
    \subcaption{Priority inheritance}
    \label{subfig:priority_inheritance}
  \end{minipage}
  \hfil
  \begin{minipage}{0.32\hsize}
    \centering
    \includegraphics[width=0.85\hsize]{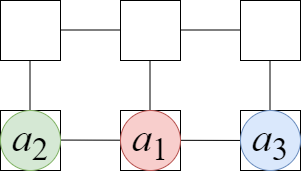}
    \subcaption{One timestep later}
    \label{subfig:1step_later}
  \end{minipage}
  \caption{Priority Inheritance in PIBT.}\label{fig:Interitance}
\end{wrapfigure}
Figure~\ref{fig:Interitance} shows an example where three agents $a_1, a_2$, and $a_3$ decide the next node by PI, where $p_1<p_2<p_3$. Figure~\ref{subfig:ikidumari} shows a situation where $a_3$ tries to move down, but $a_1$ is already there. Therefore, $a_1$ must move left to make room for $a_3$, but it is the next node of $a_2$, which has a higher priority than $a_1$, resulting in a deadlock. However, based on PI, $p_1 \leftarrow p_3$, and $a_2$ abandons its right node (Fig.~\ref{subfig:priority_inheritance}). Then, $a_1$ moves left and $a_3$ moves down (Fig.~\ref{subfig:1step_later}). Agent $a_2$ can move up or stay as it is, depending on its rank of the next nodes. Fig.~\ref{subfig:1step_later} shows the case when $a_2$ remains as is.
\par

\begin{figure}[h]
  \centering
  \begin{minipage}{0.22\hsize}
    \centering
    \includegraphics[width=0.85\textwidth]{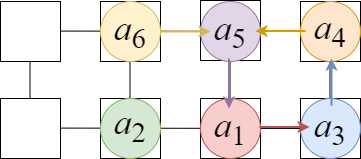}
    \subcaption{Priority inheritance (PI)}
    \label{subfig:firstpi}
  \end{minipage}
  \hfil
  \begin{minipage}{0.22\hsize}
    \centering
    \includegraphics[width=0.85\textwidth]{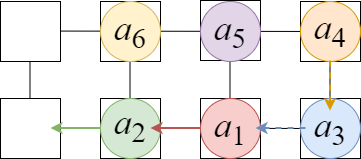}
    \subcaption{PI and Backtracking}
    \label{subfig:btandpi}
  \end{minipage}
  \hfil
  \begin{minipage}{0.22\hsize}
    \centering
    \includegraphics[width=0.85\textwidth]{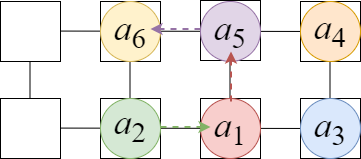}
    \subcaption{Backtracking}
    \label{subfig:backtrack}
  \end{minipage}
  \hfil
  \begin{minipage}{0.22\hsize}
    \centering
    \includegraphics[width=0.85\textwidth]{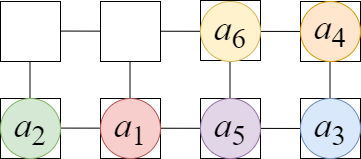}
    \subcaption{One timestep later}
    \label{subfig:later}
  \end{minipage}
\caption{Backtracking in PIBT}\label{fig:PIBTbacktrack}
\end{figure}

\begin{wrapfigure}[9]{R}[0mm]{5.3cm}
  \centering
  \includegraphics[width=0.99\hsize]{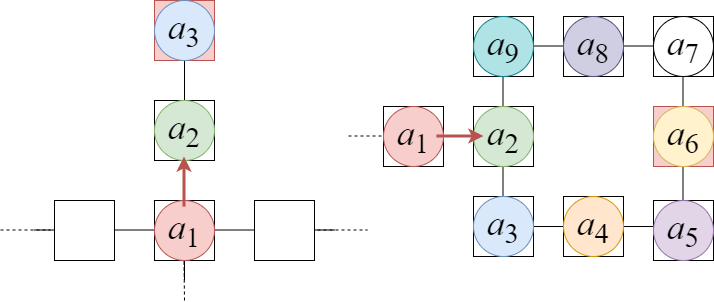}
  \caption{Examples that PIBT causes deadlocks.}
  \label{PIBT_False}
\end{wrapfigure}
When PI cascades and the involved agents are in a deadlock state, {\em backtracking} (BT) is triggered, as shown in Fig.~\ref{fig:PIBTbacktrack}, where $p_1 < p_2 < \cdots < p_6$. Fig.~\ref{subfig:firstpi} shows the situation in which $a_6$ moves right, so its priority is inherited in turn to $a_5$, $a_1$, $a_3$, and then $a_4$, but they will be deadlocked. To resolve this situation, BT conveys the occurrence of a deadlock ($\pibtFalse$) from $a_4$ in the opposite direction of PI, as shown in Fig.~\ref{subfig:btandpi}. Thus, $a_4$ and $a_3$ cancel PI, but $a_1$ can find another neighboring node to which it may be able to move because $p_1 (= p_6) > p_2$. Then, $a_1$ decides to move the node at which $a_2$ is and $a_6$'s priority is inherited to $a_2$, and $a_2$ will move left without deadlock. Agent $a_1$ conveys $\pibtTrue$ to $a_6$ in the opposite direction of PI, as shown in Fig.~\ref{subfig:backtrack}. Finally, $a_6$ moves right, but $a_3$ and $a_4$ cannot move so they stay at the current nodes (Fig.~\ref{subfig:later}).
\par

However, in a non-bi-connected environment, a deadlock occurs, as shown in Fig.~\ref{PIBT_False}, where we assume that $p_1$ is the highest. Obviously, when $a_1$ decides to move up (left figure in Fig.~\ref{PIBT_False}) or move right (right figure in Fig.~\ref{PIBT_False}), all agents are deadlocked. Furthermore, because in Okumura et al.~\cite{okumura2017}, the priority is set based on the timesteps since the agent left the starting point of its path, the priority order never changes and dead-ends cannot be resolved.

\section{Proposed Method}
\subsection{Base Algorithm}
\begin{figure}
  \hfill
  \begin{minipage}{0.20\hsize}
    \centering
    \includegraphics[width=0.95\hsize]{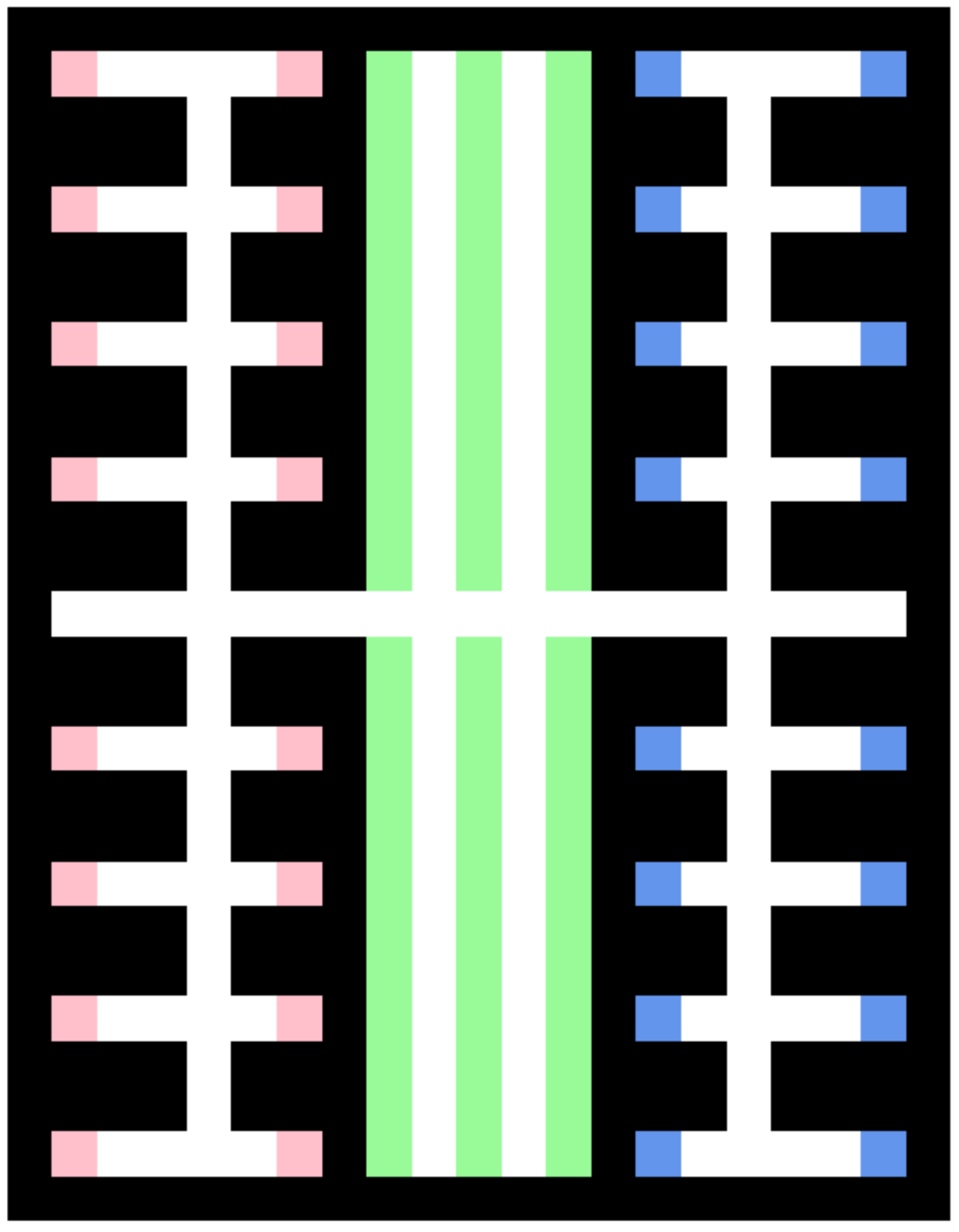}
    \subcaption{Environment 1}
    \label{env1}
  \end{minipage}\hfill
  \begin{minipage}{0.20\hsize}
    \centering
    \includegraphics[width=0.95\hsize]{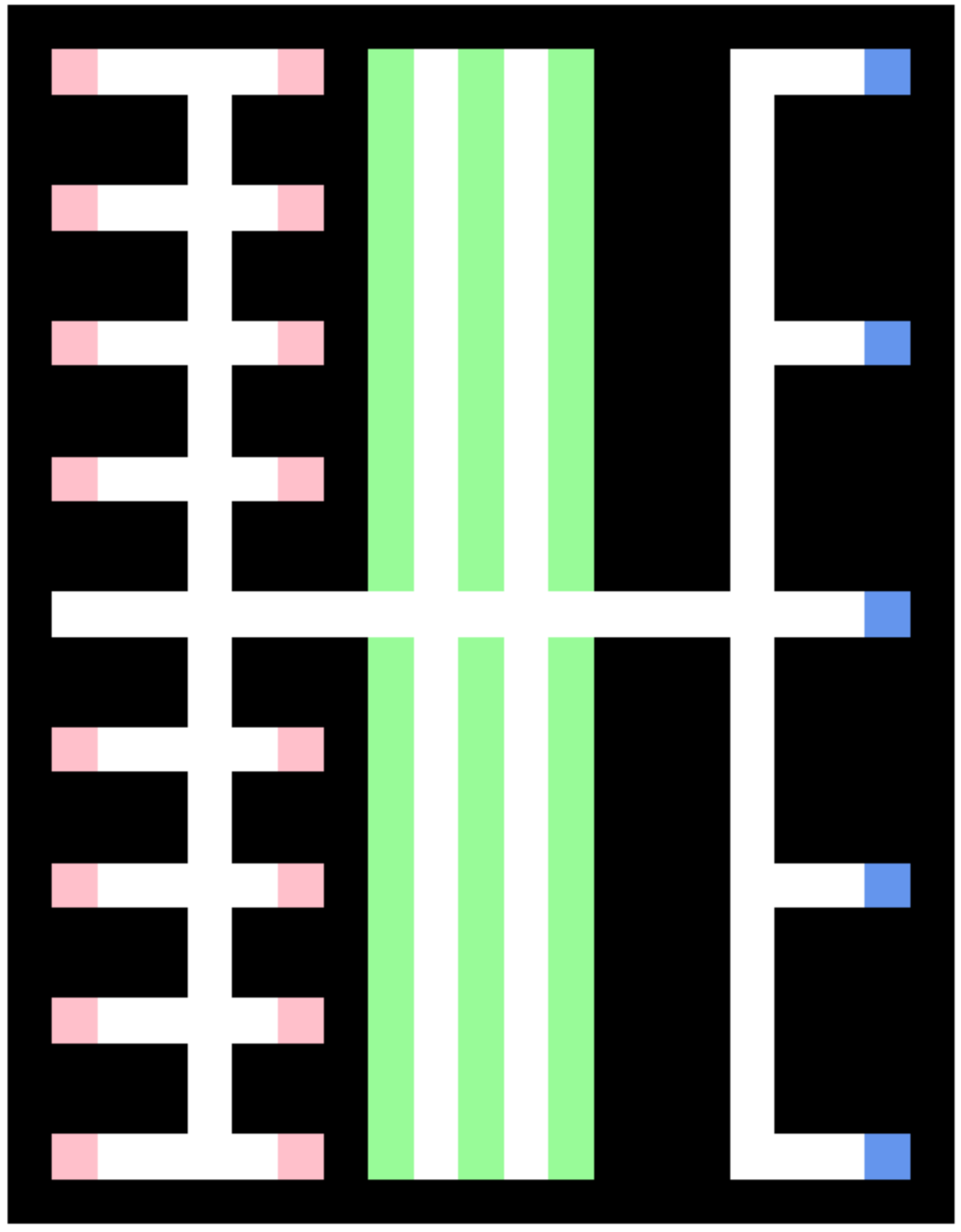}
    \subcaption{Environment 2}
    \label{env2}
  \end{minipage}\hfill
  \begin{minipage}{0.20\hsize}
    \centering
    \includegraphics[width=0.95\hsize]{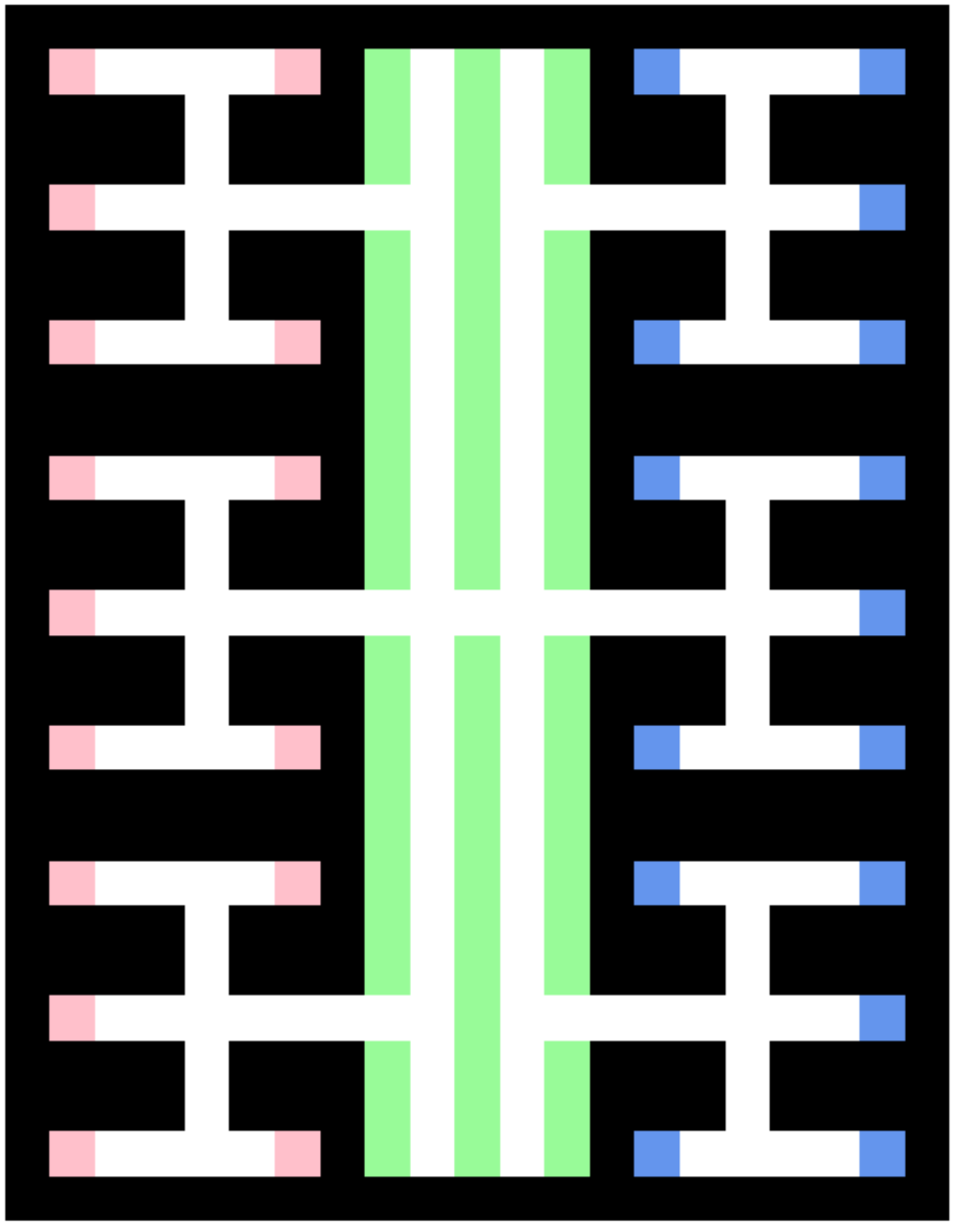}
    \subcaption{Environment 3}
    \label{env3}
  \end{minipage}\hfill
  \begin{minipage}{0.20\hsize}
    \centering
    \includegraphics[width=0.95\hsize]{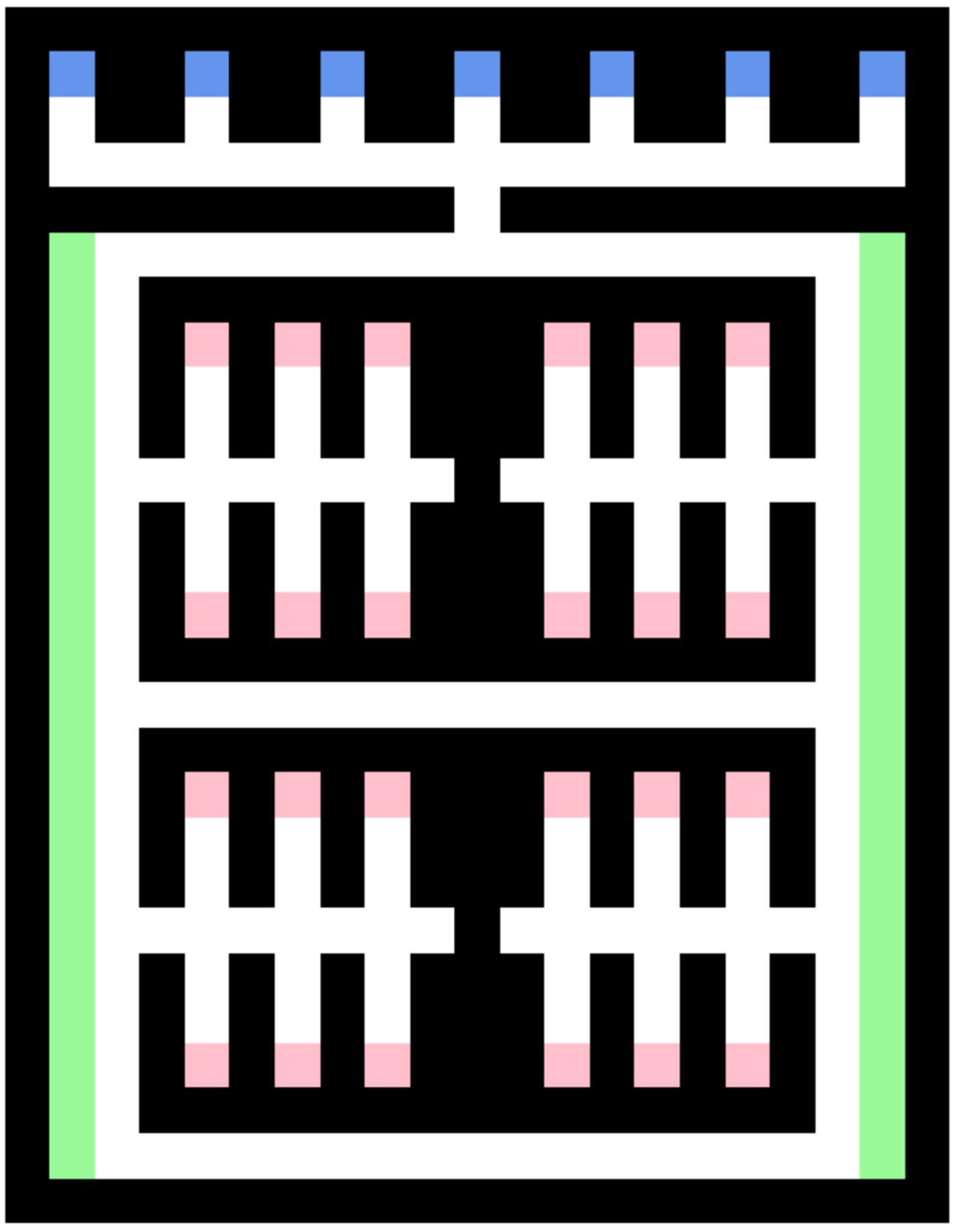}
    \subcaption{Environment 4}
    \label{env4}
  \end{minipage}
  \hfill
\caption{Example environments.}
\label{fig:environment}
\end{figure}

We propose an extension of PIBT, called {\em PIBT with Temporary Priority} (PIBTTP), to enable the continuous execution of tasks of an MAPD problem without deadlock in an environment $G=(V,E)$ consisting of a main bi-connected area with a number of tree-shaped areas, each of which is connected to a node in the main area. The subgraph describing the main area is denoted by $G_M=(V_M, E_M)$. We also denote the set of trees $\{G_T^1,\dots,G_T^B\}$, where $B$ is the number of trees and $G_T^k=(V_T^k, E_T^k)$ is the subgraph of $G$. Note that from the assumption, $V_M\cap V_T^k$ is a singleton set and its element is called the {\em connecting node}. We then define ${V'}_{-T}^{k} := V_T^k \setminus V_M\cap V_T^k$. Obviously, $V=V_M\cup {V'}_T^1\cup\cdots\cup{V'}_T^B$ is a disjoint union. Four example environments are shown in Fig.~\ref{fig:environment}, where pink and blue nodes are pickup and delivery nodes, respectively. For example, Environment~1 (Env.~1) has two trees, i.e., $G_M$ is the rectangular area shown by white and green nodes in the middle, and the trees $G_T^1$ and $G_T^2$ whose dead-ends are colored by pink or blue, respectively. Similarly, Env.~2, Env.~3, and Env.~4 have two, six, and five trees, respectively.
\par

We introduce five assumptions.
\begin{itemize}
  \item[(A1)] The number of agents is smaller than that of nodes in the main area $V_M$ as is the case with PIBT.
  \item[(A2)] For $\forall \tau=(s^\tau, g^\tau)\in\TaskSet$, $s^\tau$ and $g^\tau$ are not in the same tree.
  \item[(A3)] When an agent is in a tree, it does not choose or is not allocated a new task $\tau$ whose pickup node $s^\tau$ is in the current tree.
  \item[(A4)] An agent does not enter trees that do not include its current destination.
  \item[(A5)] When an agent is in a tree, it moves only on the shortest path to the current destination.
\end{itemize}
Note that Assumptions A2 and A3 are introduced by considering carrying tasks between, for example, storage racks/areas and loading/unloading ports of trucks in a warehouse. If agent $a_i$ in a tree is allocated the task whose pickup node is in the same tree, $a_i$ first moves to $V_M$ and then starts to execute the task to follow A3. A4 is to prevent the redundant and undesired activities. A5 will be removed to improve the efficiency in Section~\ref{sec:Efficiency}. We note again that the destination of $a_i$ allocated task $\tau$ is its pickup node $s^\tau$ or delivery node $g^\tau$.
\par

\begin{figure}
\begin{minipage}[t]{0.47\hsize}
  \begin{algorithm}[H]
  \caption{PIBTTP at time $t$.}\label{alg:PIBTTP}
    \begin{algorithmic}[1]
  \State $U \leftarrow \AgentSet$ // Agents that do not decide next nodes.\label{alg:init1}
  \State $O \leftarrow \varnothing$\; // Nodes already occupied by
  agent at $t+1$\label{alg:init2}
  \State // {\it Destination $d_i$ is the pickup or delivery node.}
  \Function {PIBTTP}{}
  \For{$a_i\in \AgentSet$}\label{alg:priority1}
  \If{$a_i$ is in a tree $T$, but $d_i$ is not in $T$}
  \State $p_i \leftarrow 1+\e_i$ // temporary priority \label{alg:priority3}
  \Else
  \State $p_i \leftarrow -f_i(v_i(t))+\e_i$ // normal priority
  \EndIf
  \EndFor \label{alg:priorityEnd}
  \While {$U\not=\varnothing$}\label{alg:CallPIBT1}
  \State {$a \leftarrow$ agent in $U$ with the highest priority}
  \State {\exPIBT($a$,$\perp$)}
  \EndWhile \label{alg:CallPIBTEnd}
  \EndFunction
  \State{}
  \Function \exPIBT{$a_i,a_j$}
  \State $U \leftarrow U \setminus \{a_i\}$
  \State {$C_i \leftarrow N_{v_i(t)} \cup \{v_i(t)\}$}\label{alg:C1}
  \If{$a_i$ is in $V_M$}\label{alg:limitS}
  \State {$C_i \leftarrow C_i \cap (\overline{O \cup \{v_j(t)\}}) \cap
    (V_M \cup {V'}_T(d_i))$}\label{alg:inMain}
  \algstore{PIBTTPstore}
  \end{algorithmic}
\end{algorithm}
\end{minipage}\hfill
\begin{minipage}[t]{0.51\hsize}
  \begin{algorithm}[H]
    \begin{algorithmic}[1]
      \algrestore{PIBTTPstore}
  \Else \; // $a_i$ is in tree ${V'}_T^k$, then
  \State{$K=\{v\in V_T^k \mid v \textrm{ is on the shortest path to }
    d_i$\}} \label{alg:inTree1}
  \State {$C_i \leftarrow C_i \cap (\overline{O \cup \{v_j(t)\}}) \cap K$}\label{alg:inTree2}
  \EndIf \label{alg:CEnd}
  \State $p_i=p_j$ // PI 
  \While {$C_i \neq \emptyset$}\label{alg:next1}
  \State $v_i^* \leftarrow \argmin_{v \in C_i} f_i(v)$
  \State $O \leftarrow O \cup \{v_i^*\}$
  \If{$\exists a_k \in U$ such that $v_i^* = v_k(t)$}\label{alg:nNext1}
  \If{\exPIBT($a_k,a_i$) is \valid}
  \State $v_i(t+1) \leftarrow v_i^*$
  \Return \valid\label{alg:nNext2}
  \Else
  \State $C_i \leftarrow C_i \setminus O$\label{alg:nNext3}
  \EndIf
  \Else \; // if there is no other agent in $v_i^*$
  \State $v_i(t+1) \leftarrow v_i^*$ \label{alg:nNext4}
  \State {// (a)} \label{alg:insert}
  \State{ \Return \valid }
  \EndIf
  \EndWhile \label{alg:nextEnd}
  \State $v_i(t+1) \leftarrow v_i(t)$ \label{alg:inv1}
  \State {\Return \invalid} \label{alg:invEnd}
  \EndFunction
  \end{algorithmic}
\end{algorithm}
\end{minipage}
\end{figure}

Algorithm~\ref{alg:PIBTTP} shows the pseudo-code of PIBTTP, which decides the next nodes to move for all agents. Before starting an MAPD instance, the system randomly initializes $\e_i$ ($i=1, \dots, n$) such that $0<\e_i<1$ and $\e_i\not=\e_j$ (if $i\not= j$). The value of $\e_i$ can be fixed until all tasks in $\TaskSet$ are completed or can be changed each time agent $a_i$ arrives at the current destination node $d_i$ of $a_i$. PIBTTP prepares two variables: $U$, which is the set of agents that have not yet decided the next nodes to which they will move, and $O$, which is the set of nodes to which agents in $\AgentSet\setminus U$ will move next [Lines \ref{alg:init1},\ref{alg:init2}]. Agents calculate their own priority [Lines \ref{alg:priority1}-\ref{alg:priorityEnd}]. The priority of agent $a_i\in \AgentSet$ is $-f_i(v_i(t))+\e_i$ if $a_i$ is in the main area $V_M$ or in the tree $V_T^k$ that contains $d_i$; thus, $p_i<0$ if $a_i$ does not arrive at its destination. However, if $\exists k_0\;$ s.t. $v_i(t)\in V_T^{k_0}$ and $d_i\not\in V_T^{k_0}$ at $t$, then $p_i=1+\e_i$ ($>1$). Note that $f_i(v)$ is the shortest path length from node $v\in V$ to $d_i$ while ignoring the presence of other agents. Next, for agent $a_i\in U$ with the highest priority, PIBTTP invokes \exPIBT$(a_i,\perp)$ [Lines~\ref{alg:CallPIBT1}-\ref{alg:CallPIBTEnd}], where in \exPIBT$(a_i,a_j)$, $a_i$ inherits the priority of $a_j$, and we set $a_j=\perp$ if there is no PI.
\par

\exPIBT($a_i,a_j$) first calculates the set $C_i$ of neighboring nodes to which $a_i$ can move [Lines \ref{alg:C1}-\ref{alg:CEnd}], where ${V'}_T(d_i)$ is the tree that includes the destination $d_i$ of $a_i$ (if $d_i\in V_M$, ${V'}_T(d_i)=\varnothing$). The nodes that $a_i$ cannot be moved to next are among the following cases:
\begin{itemize}
\item[(a)] $O\cup\{v_j(t)\}$, i.e., nodes that are already occupied or reserved by agents with higher priorities.
\item[(b)] If $a_i$ is in $V_m$, ${V'}_T^k$ that does not contain $d_i$ [Line \ref{alg:inMain}].
\item[(c)] If $a_i$ is in tree ${V'}_T^k$, nodes that are not on one of the shortest paths to $d_i$ (Assumption A5) [Lines \ref{alg:inTree1},\ref{alg:inTree2}].
\end{itemize}
\par

\begin{wrapfigure}[13]{R}[1mm]{4.5cm}
  \centering
  \includegraphics[width=0.26\textwidth]{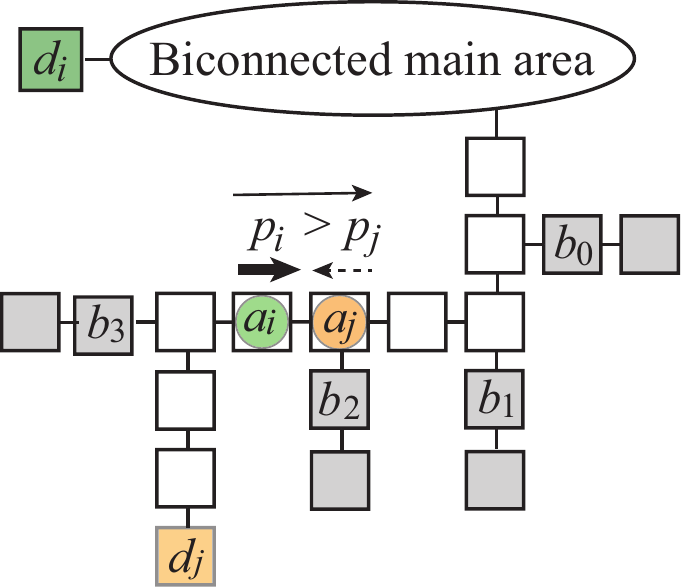}
  \caption{Encounter and then push back.}
  \label{fig:pushback}
\end{wrapfigure}
\noindent
Cases (b) and (c) mean that an agent in a tree cannot move into a side branch outside the shortest path to the current destination (see Assumption A5).
\par

Based on the rank in $C_i$ which is based on the distance $f_i(v)$ ($v\in C_i$) to $d_i$ (a smaller value implies a higher ranking), \exPIBT decides the next node to be moved (this part is almost identical to the original PIBT) [Lines~\ref{alg:next1}-\ref{alg:invEnd}]. If another agent $a_k$ is at the highest ranking node $v_i^*$ of $a_i$, \exPIBT($a_k,a_i$) is called. Then, if \exPIBT($a_k,a_i$) returns $\valid$, i.e., $a_k$ can decide the next non-conflict node to move, $a_i$ decides to move to $v_i^*$ next and \exPIBT($a_i,a_j$) returns $\valid$ [Lines \ref{alg:nNext1}--\ref{alg:nNext2}]. However, if \PIBT($a_k,a_i$) returns $\invalid$, $a_i$ cannot move to $v_i^*$ and so removes $v_i^*$ from $C_i$ [Line~\ref{alg:nNext3}]. In contrast, if there is no agent in $v_i^*$, $a_i$ decides to move to $v_i^*$ and \exPIBT($a_i,a_j$) returns $\valid$ [Line~\ref{alg:nNext4}]. Finally, if $C_i=\varnothing$, $a_i$ remains the current node and \exPIBT($a_i,a_j$) returns $\invalid$ [Lines~\ref{alg:inv1}-\ref{alg:invEnd}].
\par

Temporary priority ([Line \ref{alg:priority3}]) prevents deadlock in a tree. An example is shown in Fig.~\ref{fig:pushback}, where white nodes are on the shortest path to $d_j$ and two agents $a_i$, which moves to $d_i$ (see Assumption A3) with the temporary priority after it reached its destination at a dead-end in this tree, and $a_j$, which heads to $d_j$, are encountered (so $p_i > p_j$). Because $a_i$ has higher priority (bold arrow) and both agents cannot enter branches (Assumption A5), $a_i$ will push back $a_j$ to the bi-connected main area along the white nodes, which is the shared nodes in the shortest paths to their destinations; then, $a_i$ returns to the normal priority when it arrives at the connecting node in the main area. Thus, the following lemma is trivial.
\begin{lemma}\label{lemma1}
  Agent that has reached its destination on a tree can then reach the main area.
\end{lemma}
\noindent
Furthermore, because both the task and the agents inside the tree are finite, an agent heading for a destination in the tree can be pushed back at most a finite number of times. Therefore,
\begin{lemma}\label{lemma2}
  If an agent with a destination inside a tree reaches the connecting node of that tree in the main area, it can reach its destination as well. 
\end{lemma}
\noindent
Note that an agent heading for its destination in a tree could be pushed back to the main area. This would be inefficient, and will be discussed in Section~\ref{sec:Efficiency}.
\par

Thus, even if an agent with a new destination is inside a certain tree, it can always reach the main area (Lemma~\ref{lemma1}) and is never pushed back to the previous tree from which it has escaped (Assumption A4). Then, if its destination is inside the main area, reachability to the destination is guaranteed by original PIBT. If the destination is inside another tree, PIBT guarantees the reachability of the connecting node of that tree in the main area, and then it can reach the destination (Lemma~\ref{lemma2}). Therefore, we can obtain the following reachability.
\begin{theorem}\label{theorem1}
  Let environment $G$ consist of a bi-connected main area and several trees, each of which is extending from a node in the main area. If the set of pickup and delivery tasks $\TaskSet$ is finite, agents can complete all tasks in $\TaskSet$ within a finite number of timesteps.
\end{theorem}
\par

\subsection{Improvement for Efficient Movement in Tree}\label{sec:Efficiency}
In PIBTTP, agents with lower priority may have to return to the main area by being pushed back by the agent that has reached its destination inside the tree, and therefore has temporarily higher priority. Because this is quite inefficient, we attempt to extend PIBTTP so that agents can avoid to a side path (branch) that exists along the way when being pushed back to the main area. This extended algorithm is called the {\em PIBTTP with Temporary Avoidance} (PIBTTP-TA).

\begin{figure}
  \begin{minipage}[t]{0.48\hsize}
    \begin{algorithm}[H]
      \caption{PIBTTP-TA}\label{alg2:PIBTTP-TA}
      \begin{algorithmic}[1]
        \State{$U, O$ and $d_i$ are identical in PIBTTP}
        \State{$R=\varnothing$: reserved nodes}
        \State{$V_{TA}$: nodes in which agents in TAS exist.}
        \Function {PIBTTP-TA}{}
        \For{$a_i\in \AgentSet$}
        \If{$v_i(t)\in {V'}_T(d_i)$}
        \If{$a_i$ is in TAS}\label{alg2:priority1}
        \State $p_i \leftarrow \e_i$ // temporary priority \label{alg2:priority2}
        \EndIf 
        \ElsIf{$a_i$ is in a tree $V_T^{k_0}\not= {V'}_T(d_i)$}
        \State $p_i \leftarrow 1+\e_i$ // temporary priority
        \Else
        \State $p_i \leftarrow -f_i(v_i(t))+\e_i$ // normal priority
        \EndIf
        \EndFor \label{alg2:priorityEnd}
        \While {$U\not=\varnothing$}\label{alg2:CallPIBT1}
        \State {$a \leftarrow$ agent in $U$ with the highest priority}
        \State {\exPIBTTA($a$,$\perp$)}
        \EndWhile \label{alg2:CallPIBTEnd}
        \EndFunction
      \end{algorithmic}
    \end{algorithm}
  \end{minipage}
  \hfill
  \begin{minipage}[t]{0.48\hsize}
    \begin{algorithm}[H]
    \caption{Function \exPIBTTA($a_i$,$a_j$)}\label{alg2:exPIBT-TA}
    \begin{algorithmic}[1]
      \State{Part I: Replacing the \ref{alg:limitS}nd to \ref {alg:CEnd}th lines
        in \exPIBT with the following lines.}
      \If{$a_i$ is in $V_M$}
      \State {$C_i \leftarrow C_i \cap (\overline{O \cup \{v_j(t)\}}) \cap
        (V_M \cup {V'}_T(d_i))$}
      \ElsIf{$a_j\not= \perp$} // i.e., $p_i$ is inherited from $a_j$ \label{alg2:restrict1}
      \State {$C_i \leftarrow C_i \setminus (O \cup \{v_j(t)\}\cup
        R\cup V_{TA})$} \label{alg2:restrict2}
      \Else
      \State{$K=\{v\in V_T^k \mid v \textrm{ is on a shortest path to }
        d_i$\}} // (b) \label{alg2:inTree1}
      \State{$C_i \leftarrow C_i\setminus (O \cup \{v_j(t)\}\cup
        \overline{K})$} \label{alg2:pushed}
      \EndIf
      \State{}
      \State{Part II: Insert the following lines to
        Line~\ref{alg:insert} in \exPIBT.}
      \If{$a_i$ is in ${V'}_T(d_i)$ and $v_i^*\in\overline{K}$}
      \State $a_i$ is in TAS; \Reserve($a_i$) \label{alg2:reserve}
      \Else
      \State $a_i$ is reverted from the TAS; \Revert($a_i$) \label{alg2:dereserve}
      \EndIf
    \end{algorithmic}
  \end{algorithm}
\end{minipage}
\end{figure}

Suppose that agent $a_j$ in ${V'}_T(d_j)$ is pushed back toward the connecting node of ${V'}_T(d_j)$ by a higher priority agent $a_i$, as shown in Fig.~\ref{fig:pushback}. To achieve the temporary avoidance of agent $a_j$ in the way of $a_i$, $a_j$ is not directed to the connecting node along the white nodes, but preferentially to an entrance node of another branch ($b_2$ in Fig.~\ref{fig:pushback}) if possible. Note that this entrance node is located next to a node on the shortest path. At this time, this algorithm adds the node to which agent $a_j$ should originally proceed toward its destination (the node located at $a_i$ in Fig.~\ref{fig:pushback}) into $R$ as a reserved node, sets $a_j$ to the {\em temporary avoiding state} (TAS), and raises the priority of the waiting $a_j$ to $\e_j$ ($>0$). Moreover, we assume that another agent $a_k$ from the end of the tree whose destination $d_k$ is beyond the reserved node can pass through the reserved node because $a_k$ has higher priority $1+e_k$; however, other agents are not allowed to pass through, so $a_j$ is not pushed to an inner node of the branch. This reservation information is shared with agents in the same tree.
\par

We show the pseudo-code for PIBTTP-TA in Algorithms~\ref{alg2:PIBTTP-TA} and \ref{alg2:exPIBT-TA}. We only describe the differences between \exPIBTTA from \exPIBT in Algorithm~\ref{alg2:exPIBT-TA}. When agent $a_i$ is in the TAS, its priority is set to $\e_i$ ($0<\e_i<1$) [Lines~\ref{alg2:priority1} and \ref{alg2:priority2} in Algorithm~\ref{alg2:PIBTTP-TA}]. Then, function \exPIBTTA is invoked. There are two differences between \exPIBTTA and \exPIBT. First, the set $C_i$ of nodes to which $a_i$ can move next is modified. In particular, when the priority of $a_i$ has been inherited from another agent, $C_i$ includes neighboring nodes outside of the shortest path to $d_j$ for temporary avoidance [Lines~\ref{alg2:restrict1}, \ref{alg2:restrict2} in Algorithm~\ref{alg2:exPIBT-TA}]. The second difference is that when $a_i \in {V'}_T(d_i)$ and $v_i^*\not\in K$, $a_i$ is set in the TAS and calls \Reserve($a_i$), where $a_i$ enters the TAS, node $v_o$, which should have been the next node in order for $a_i$ to move toward the destination, is added to $R$. Otherwise, $a_i$ reverts from the TAS and excludes node $v_o$ that it had reserved from $R$ only if no other agent has reserved it.
\par

It is clear that Lemma~\ref{lemma1} holds for algorithm PIBTTP-TA
because the agent that has arrived at the destination in a tree has a
high priority. Second, agent $a_i$ in the TAS can return to the shortest path to $d_i$ in the current tree because its priority is positive ($p_i>0$) and it is in the node next to the branch point of the shortest path. Moreover, it is never pushed to an inner node of the branch [Line~\ref{alg2:pushed}]. Thus, it can return to the shortest path at some time and follow that path to its destination. This indicates that Lemma~\ref{lemma2} also holds for PIBTTP-TA. Thus, we can obtain the same result:
\begin{theorem}\label{theorem2}
  Under the same condition of Theorem~\ref{theorem1}, agents can complete all tasks in $\TaskSet$ with PIBTTP-TA.
\end{theorem}

\section{Experimental Evaluation}
\subsection{Experimental Setting}
To evaluate our proposed method, we conducted the experiments in the environments shown in Fig.~\ref{fig:environment}. We set TP~\cite{ma2017} as the baseline method because it is a well-known algorithm for MAPD. Agents are initially placed on green nodes and start to perform an MAPD instance, which consists of 50 tasks ($|\TaskSet|=50$). Note again that pink and blue nodes in these environments are pickup and delivery nodes, respectively. Env.~1 assumes a warehouse in which the pickup and delivery nodes are located at dead-ends of deep trees, i.e., trees that have relatively large depths. This environment is advantageous for TP because there are many and the same number of pickup and delivery nodes, and it is relatively disadvantageous for PIBTTP and PIBTTP-TA because there is a possibility of pushback to the connecting node over long distances in a deep tree. We are interested in determining whether PIBTTP-TA can increase efficiency in this environment. Env.~2 has one deep tree and an unbalanced number of pickup and delivery nodes. Env.~3 is similar to Env.~1, but we have reduced the depth of the trees on the left and right sides. Env.~4 was designed to be more similar to a real warehouse environment; the materials are stored in densely arranged racks located in the middle of the environment, the access paths to them are tree-like structures, and agents must deliver the materials to the blue nodes in an upper tree for loading on trucks. Generally, the number of storage racks is large and the number of points loading trucks is limited, resulting in an imbalance between the two types of nodes.
\par

Fifty tasks in $\TaskSet$ are generated initially by selecting pickup and delivery nodes in each environment. When an agent finishes the current task, it randomly selects a new task from $\TaskSet$ and continues to perform it until all tasks are completed. However, because TP has restrictions on the tasks that can be selected~\cite{ma2017}, agents with TP choose tasks from $\TaskSet$ to meet the requirements. When $\TaskSet$ becomes empty or when agents cannot select tasks from $\TaskSet$ owing to the restriction, they return to their initial nodes so that they do not obstruct other agents. The number of agents $n$ was varied from 5 to 40 in increments of 5. The data presented below is the average time to complete all 50 tasks in the taskset for over 200 trials.
\par

\begin{figure}
  \begin{minipage}{0.23\hsize}
    \centering
    \includegraphics[width=0.99\hsize]{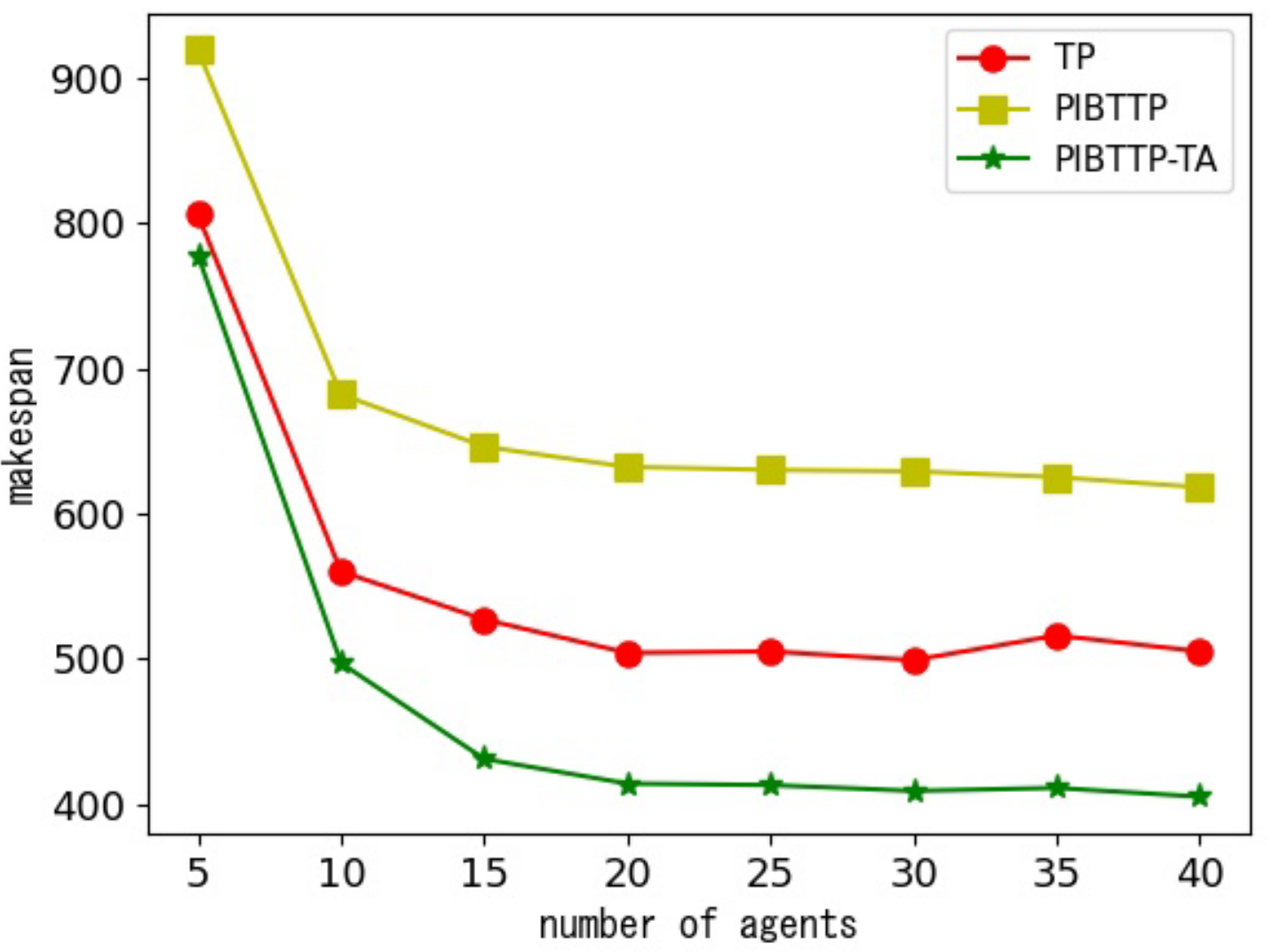}
    \subcaption{Env.~1}
    \label{subfig:env1}
  \end{minipage}\hfill
  \begin{minipage}{0.23\hsize}
    \centering
    \includegraphics[width=0.99\hsize]{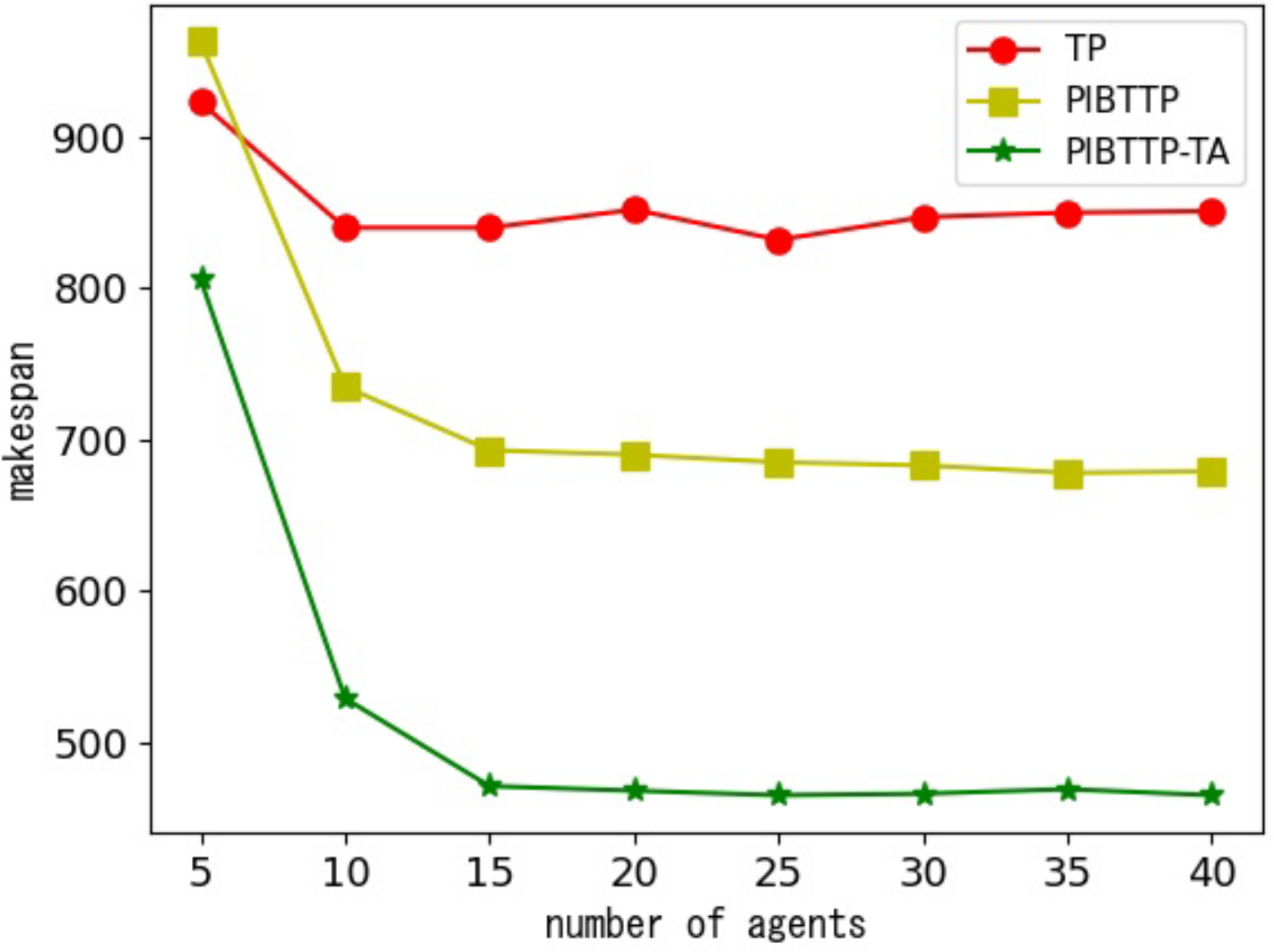}
    \subcaption{Env.~2}
    \label{subfig:env2}
  \end{minipage}\hfill
  \begin{minipage}{0.23\hsize}
    \centering
    \includegraphics[width=0.99\hsize]{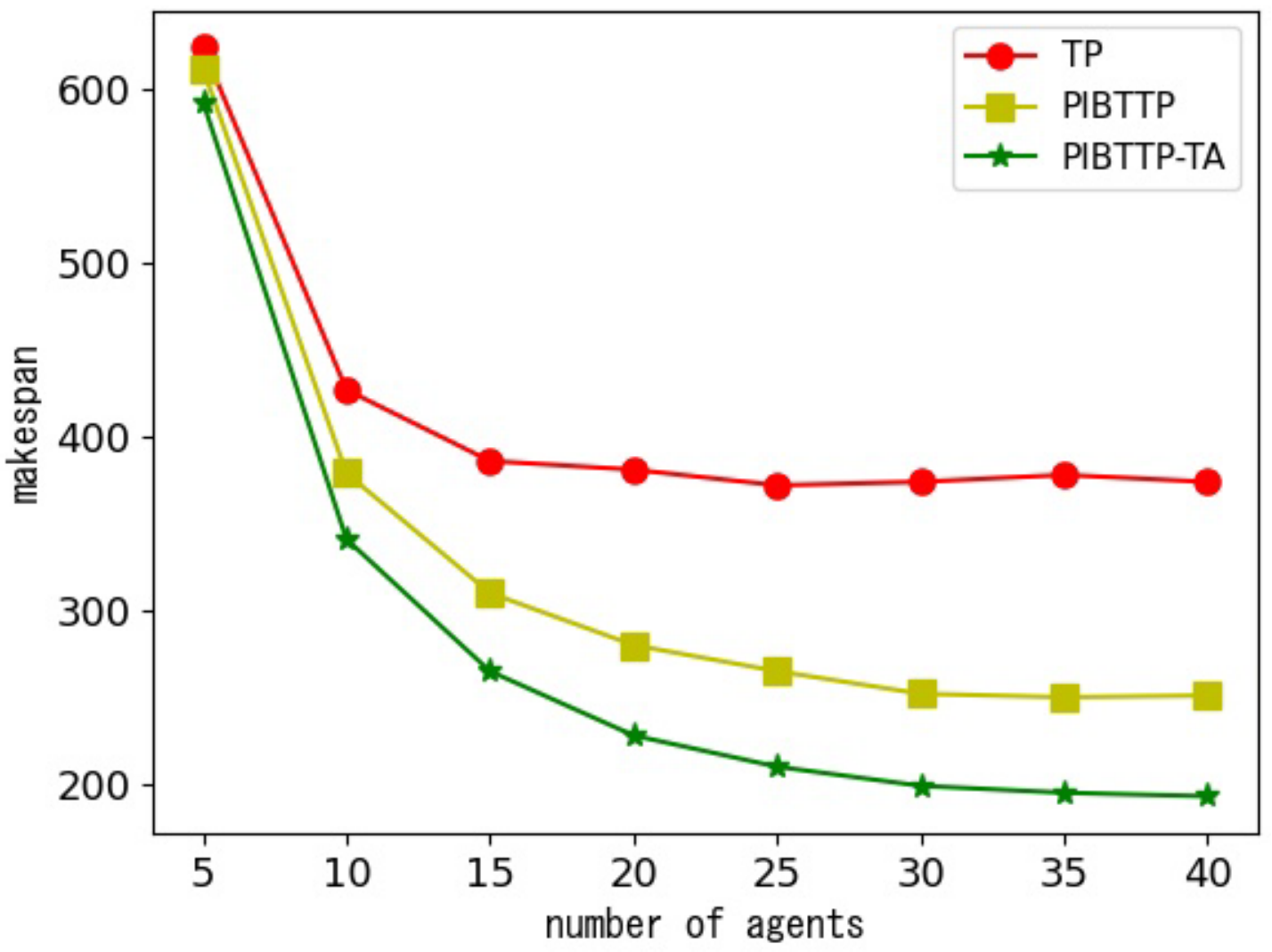}
    \subcaption{Env.~3}
    \label{subfig:env3}
  \end{minipage}\hfill
  \begin{minipage}{0.23\hsize}
    \centering
    \includegraphics[width=0.99\hsize]{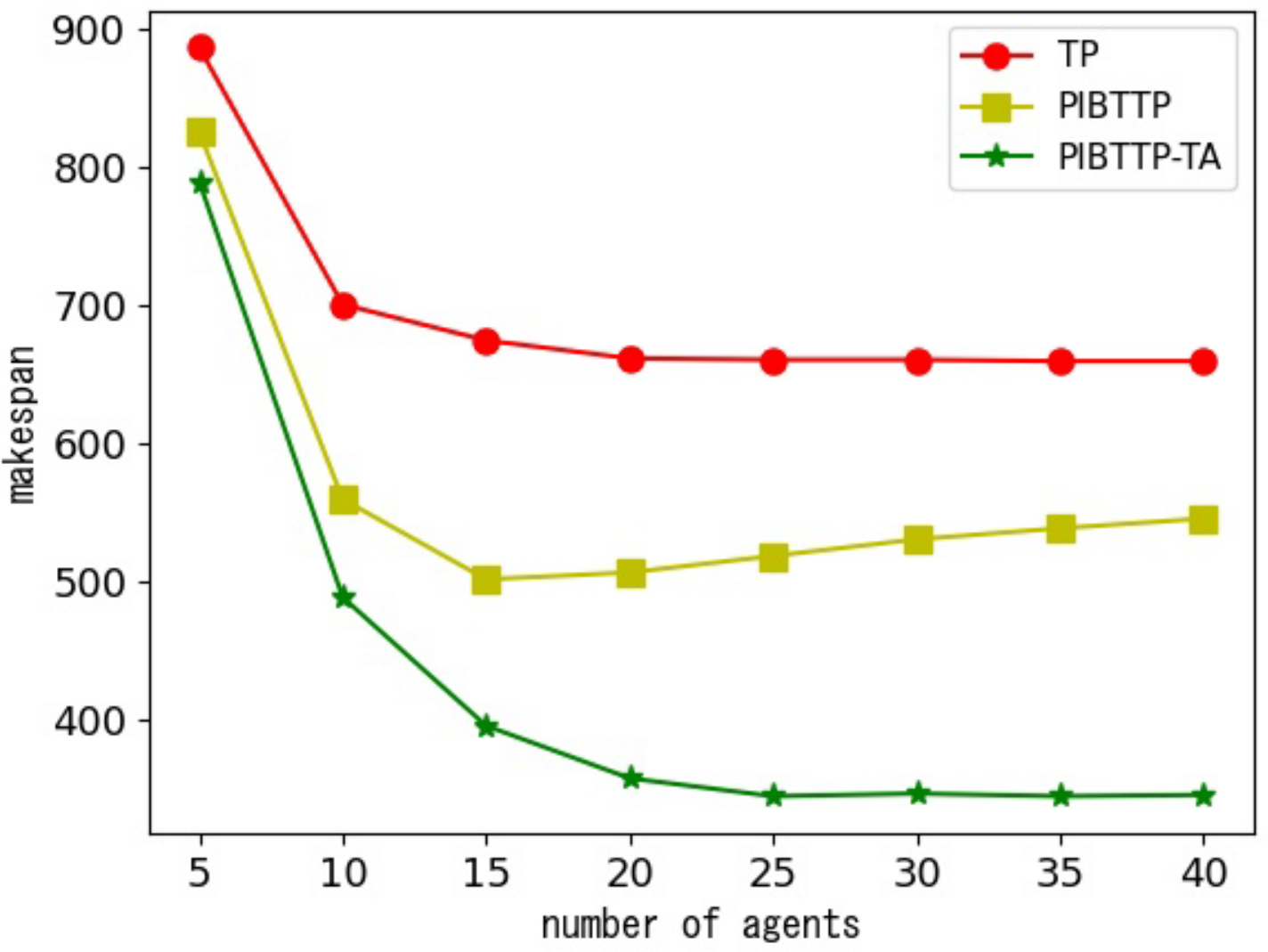}
    \subcaption{Env.~4}
    \label{subfig:env4}
  \end{minipage}
\caption{Makespan --- Experimental Results.}
\label{fig:results}
\end{figure}

\subsection{Experimental Results and Discussion}
The {\em makespan}, i.e., the time required to complete all tasks in each environment, is shown in Fig.~\ref{fig:results}. First, if we compare the results of TP and PIBTTP, we can see that the efficient methods differed depending on the environment. For example, Fig.~\ref{subfig:env1} shows that in Env.~1, the makespan of PIBTTP is large and so its efficiency is lower than that of TP, while Fig.~\ref{subfig:env2} shows that when $n \geq 10$, the makespan of PIBTTP is smaller and PIBTTP is more efficient than TP in Env.~2. This may be because of the parallelism and overhead caused by the restrictions of each algorithm. In TP, each agent must select a task so that no more than two agents simultaneously aim at the same destination, i.e., pickup and delivery nodes. Env.~1 has 16 pickup and delivery nodes each, allowing 16 agents to move simultaneously. However, Env.~2 has only five delivery nodes, and therefore, when $n\leq 5$, all agents can perform tasks simultaneously, but no further parallelism is possible.
\par

In contrast, because PIBTTP does not have the restrictions as required by TP, multiple agents can simultaneously select tasks with the same pickup/delivery nodes, and the efficiency in Env.~2 thus became high because of the high parallelism. Many agents with PIBTTP could also move around in Env.~1 simultaneously. However, because this environment has two deep trees whose depths are high and which have many pickup/delivery nodes, agents heading to destinations concentrate on these trees and may frequently be pushed back to their connecting nodes in the main area, incurring a high overhead.
\par

Moreover, PIBTTP also outperforms TP in Envs.~3 and 4, as shown in Figs.~\ref{subfig:env3} and \ref{subfig:env4}. This is because Env.~3 has the same number of pickup and delivery nodes, but the trees are shallower, so agents could move in and out of individual trees in a short time, reducing the concentration of agents in the same tree. Env.~4 also has slightly deeper trees and unbalanced numbers of pickup and delivery nodes, and thus it is advantageous for PIBTTP. We consider closely Fig.~\ref{subfig:env4}, which shows that the efficiency of PIBTTP is slightly reduced with an increase in the number of agents $n$ when $n\geq 20$. This may be due to the size and the shape of the main area. In particular, regions around the nodes in the main area that is connected to trees in Env.~4 (Fig.~\ref{env4}) are narrower than those in other environments. Thus, as the number of agents increased, more agents were temporarily pushed back into the main area, which may be the main reason for the loss of efficiency because such agents prevented other agents from moving in the main area.
\par

In contrast to PIBTTP, the improved algorithm, PIBTTP-TA, always outperformed TP (and PIBTTP), as shown in Fig.~\ref{fig:results}. As shown in Figs.~\ref{env1} and \ref{env2}, these environments have deep trees with several branches. Thus, an agent can temporarily avoid one of the branches if possible, and could prevent much of the overhead from being pushed back to the connecting nodes of the trees. In Env.~3, the difference in the efficiency between PIBTTP and PIBTTP-TA was smaller than in other environments because the depths of trees in Env.~3 were small and the overhead of agents being pushed back to the main area was relatively small. In Env.~4, the efficiency decreased with an increase in the number of agents when $n\geq 15$ in PIBTTP, while the efficiency increased in PIBTTP-TA. This may be because agents with PIBTTP-TA avoid branches within the trees and PIBTTP-TA could reduce crowding in the narrow region in the main area near the connecting nodes of the trees.
\par

\subsection{Discussion}
The experimental results indicate that PIBTTP and PIBTTP-TA improved parallelism compared to TP. The improvement is especially significant when the number of pickup nodes and delivery nodes are much different. Furthermore, PIBTTP-TA could make movements more efficient in the tree region that is added to extend PIBT in this proposal, and it could achieve a more efficient behavior than TP even in environments where the number of pickup and delivery nodes are almost equal, and so appears to benefit TP. Note that all of the experimental environments do not meet the conditions required by PIBT. However, PIBTTP and PIBTTP-TA may cause deadlock in the environment with loops, as shown in the right side of Fig.~\ref{PIBT_False}. We will address this limitation as our future work.

\section{Conclusion}
For the MAPD problem, we proposed PIBTTP, which is an extension of the existing PIBT, and which can be used for environments where the conditions required by PIBT are relaxed to enhance its applicability, and PIBTTP-TA, which is the more efficient version of PIBTTP. We experimentally demonstrated that by using temporary priority and limiting the direction of movement, multiple agents with PIBTTP can continuously carry materials cooperatively without deadlocking in environments with relaxed conditions. The proposed method, PIBTTP, has high concurrency and is considerably more efficient than TP, which was used as a baseline in our experiments, and is often used as a comparative method in many studies. However, PIBTTP has a disadvantage in that the depths of trees connected to the bi-connected main area significantly impact their efficiency. Thus, we proposed the further improved PIBTTP-TA to eliminate this disadvantage to some extent, and we showed that it could achieve high efficiency.
\par

Future work includes proposing a method that allows for the continuous execution of the MAPD task even in environments with loops, and more generally in environments in which multiple bi-connected areas are connected.

\end{document}